\documentclass[aps,prc,preprint,tightenlines,showpacs,%
amssymb,byrevtex,nofootinbib]{revtex4}  %endfloats

\usepackage{epsfig,bm,dcolumn}
\newcommand{\fslash}[1]{\ooalign{\hfil/\hfil\crcr$#1$}}
\begin{document}

%\preprint{hep-ph/}

%%%%%%%%%%%%%%%%%%%%% Title %%%%%%%%%%%%%%%%%%%%%%

\title{ Parity of $\Theta^+$(1540) from QCD sum rules}

%%%%%%%%%%%%%%%%%%%% Authors %%%%%%%%%%%%%%%%%%%%%
%%%%%%%%%%%%%%%%%%%% Addresses %%%%%%%%%%%%%%%%%%%%%

\author{Su Houng Lee}%
\email{suhoung@phya.yonsei.ac.kr}

\author{Hungchong Kim}%
\email{hungchon@postech.ac.kr}
\altaffiliation[Present address: ]{Department of Physics,
Pohang University of Science and Technology, Pohang 790-784 Korea
}

\author{Youngshin Kwon}%
\email{kyshin@yonsei.ac.kr} 

\affiliation{Institute of Physics and Applied Physics,
Yonsei University, Seoul 120-749, Korea}

%\date{\today}

%%%%%%%%%%%%%%%%%%%% Abstract %%%%%%%%%%%%%%%%%%%%%

\begin{abstract}

The QCD sum rule for the pentaquark $\Theta^+$, first
analyzed by Sugiyama, Doi and Oka, is reanalyzed with a
phenomenological side that explicitly includes the contribution
from the two-particle reducible kaon-nucleon intermediate state.
The magnitude for the overlap of the $\Theta^+$ interpolating
current with the kaon-nucleon state is obtained by using soft-kaon
theorem and a separate sum rule for the ground state nucleon with
the pentaquark nucleon interpolating current.
It is found that the K-N intermediate state constitutes only 10\%
of the sum rule so that the original claim that the parity of
$\Theta^+$ is negative remains valid.

\end{abstract}


\maketitle

\section{Introduction}

The discovery of the $\Theta^+$ baryon by LEPS Collaboration at
SPring-8 \cite{LEPS03} and subsequent confirmation
in the other experiments~\cite{OTHERS}
have spurred a lot of works in the field of
exotic hadrons.   So far, not much is known about the properties
of the $\Theta^+$ except its mass, which is about 1540 MeV, and
its small decay width, which is smaller than the experimental
resolutions of around $10$ MeV. Thus, to determine the
quantum numbers as well as  other properties of $\Theta^+$,
various production
mechanisms have been proposed~\cite{reactions}.
The $\Theta^+$ baryon, being a
strangeness +1 state, is exotic since its minimal quark content
should be $uudd\bar{s}$.
Other states that have positive
strangeness but different charges are not
observed~\footnote{For various classifications of the pentaquarks
and their decay modes, see Ref.\cite{class}.},
which suggests
that the $\Theta^+$ is an isosinglet.
The existence of such an
exotic state with narrow width and spin-parity $1/2^+$ was first
predicted by Diakonov {\it et al.\/} \cite{DPP97} in the chiral
soliton model, where the $\Theta^+$ is a member of the baryon
anti-decuplet.
The positive parity is also supported by
the constituent quark model with flavor-spin hyperfine 
interaction~\cite{Stancu:2003if}, 
the diquark-diquark-antiquark
picture of Jaffe and Wilczek (JW) \cite{JW03},the triquark-diquark picture
\cite{KL03a},
the quark potential model calculations \cite{CCKN03b},
and the constituent quark model where chiral dynamics are
included\cite{Hosaka03}.
On the other hand,
it is expected in a naive constituent quark model that the ground state of the
pentaquark have a negative parity because all the quarks would
be in the $s$-state. The negative parity is supported by
the calculations based on QCD, such as
the lattice calculation~\cite{lattice} and
QCD sum rules~\cite{Oka1}.
Hence, determining the parity of the pentaquark states will not only be
important in establishing the basic quantum numbers of the
pentaquark states, but also in understanding the QCD dynamics
especially when multiquarks are involved.

Currently the results from both lattice QCD\cite{lattice} and QCD
sum rule\cite{Oka1} analysis, which show the existence of a
negative parity pentaquark state in the isospin zero and spin 1/2
channel, face a challenge to be settled.
In particular, subsequent analysis in the lattice QCD found no stable
pentaquark state in the advertised channel\cite{Chiu:2004gg}.
Similarly, in a different QCD sum rule analysis, the parity was
found to be positive~\cite{Kondo:2004cr}.
A major uncertainty in both approaches is
associated with isolating the pentaquark contribution in the
correlation functions between the pentaquark interpolating
currents. Because the interpolating current can also couple to the
two-hadron reducible (2HR) kaon-nucleon (K-N) intermediate
state, it is difficult to extract signals for the
pentaquark state from the theoretical calculation of
the two point correlation function. This is particularly so
because the K-N threshold lies below the expected $\Theta^+$ state
and neither the Borel transformation in the QCD sum rule nor the large
imaginary
time behavior in the lattice calculation can isolate
the $\Theta^+$ state. Hence, it is essential in both approaches to
estimate the contribution coming from the K-N intermediate state
in the correlation function.

This point has been noted for the QCD sum rule approach by
Kondo {\it et al.}\cite{Kondo:2004cr}, who claimed that after subtracting
out the 2HR part of the operator product expansion(OPE), one finds
that the parity becomes positive.
However, as we will show, subtracting the 2HR contribution in the
OPE level is an ill-defined approach.  Instead, its contribution
can be estimated in the phenomenological side. The magnitude for
the overlap of the $\Theta^+$ interpolating current with the
kaon-nucleon state is obtained by first applying a chiral
rotation to the $\Theta^+$ interpolating current and estimate the
kaon overlap in the soft-kaon limit. We then analyze the QCD sum
rule for the nucleon with the resulting pentaquark nucleon
interpolating current to estimate the nucleon overlap. It is found
that the K-N intermediate state constitutes only 10\% of the sum
rule so that the original claim that the parity of $\Theta^+$ is
negative remains valid.

This paper is organized as follows.
In section II, we present our method of treating the K-N intermediate
state appearing in the correlation function for the $\Theta^+$ sum rule.
We then calculate the overlap of the $\Theta^+$ interpolating field
with the K-N intermediate state in Section III.
In section IV, we reanalyze the $\Theta^+$ sum rule  after subtracting
out the K-N intermediate state.

\section{Correlation function}

Let us begin with the correlation function between the
interpolating field for $\Theta^+$,
\begin{eqnarray}
\Pi(q)=i \int d^4 x e^{iq\cdot x} \langle 0 | T [J_\Theta(x),
\bar{J}_\Theta (0)] |0 \rangle \label{corr}
\end{eqnarray}
where
\begin{eqnarray}
J_\Theta &  = & \epsilon^{abc} \epsilon^{def} \epsilon^{cfg} \{
u^T_a C d_b \} \{ u^T_d C \gamma_5 d_e \}  C\bar{s}^T_g \ ,
\nonumber \\
\bar{J}_\Theta &= &  - \epsilon^{abc} \epsilon^{def}
\epsilon^{cfg} \{ \bar{d}_e \gamma_5 C \bar{u}^T_d \} \{ \bar{d}_b
C \bar{u}^T_a \} s^T_g C\ .
\end{eqnarray}
Here the roman indices denote the color, $C$ the charge conjugation
and the superscript $T$ transpose.
The OPE of this correlation function has been calculated by
Sugiyama, Doi and Oka (SDO)\cite{Oka1} 
and its extension to the anti-charmed
pentaquark has been made in Ref.\cite{Kim:2004pu}. From comparing the OPE
to the phenomenological side saturated by the ground state
$\Theta^+$ and a continuum, SDO were able to identify the parity of
the $\Theta^+$ to be negative.

However, as has been noted by Kondo, Morimatsu and Nishikawa
(KMN)\cite{Kondo:2004cr}, the correlation function
can have two-hadron reducible (2HR)
contributions in addition to the two-hadron irreducible (2HI) part.
This means that since $J_\Theta$ is the 5 quark current with a strangeness +1,
isospin zero, it can also easily couple directly
to a kaon-nucleon intermediate state or any of their excited
states, namely,
\begin{eqnarray}
\Pi(q) & = & \Pi^{2HI}+ \Pi^{2HR}
\end{eqnarray}
where
\begin{eqnarray}
\Pi^{2HI} & = & - \frac{ |\lambda_\Theta|^2} { \fslash{q}-m_\Theta}
\cdot\cdot\cdot  \nonumber \\
\Pi^{2HR} & = & -i \int \frac{ d^4p} {(2\pi)^4}  \Pi_N(p)
\Pi_{K}(p-q) \ .
\end{eqnarray}
Therefore, to extract information about the pentaquarks from the
OPE calculation, one has to subtract the contributions from the
2HR contributions,
\begin{eqnarray}
\Pi^{2HI}(q)& =& \Pi^{OPE}(q)-\Pi^{2HR}(q) \nonumber \\
& = & \Pi^{OPE}(q)   + i \int \frac{ d^4p} {(2\pi)^4} \Pi_N(p)
\Pi_{K}(p-q)\ . \label{2hi}
\end{eqnarray}
In the left hand side (LHS), we are interested in the ground state
but the OPE part can be calculated for large $-q^2$.  The
usual way of matching the two sides with different regions of the momentum
is achieved by the Borel transformation.
A question in this particular case is how to subtract the 2HR
contribution effectively.

\subsection{Method by KMN}

KMN suggest to calculate the large $-q^2$ limit of $\Pi^{2HR}(q)$
using the OPE of the $\Pi_N^{OPE}(p), \Pi_{K}^{OPE}(p-q)$, namely
\begin{eqnarray}
\Pi^{2HR,OPE}(q)  = -i \int \frac{d^4p}{(2 \pi)^4} \Pi_N^{OPE}(p)
\Pi_{K}^{OPE}(p-q)\ . \label{KMN}
\end{eqnarray}
Under this prescription,  KMN found that the contribution from the 2HR is
large enough to change the previous result on the $\Theta^+$ parity.
However, a little inspection shows that such factorization is
ill-defined.  The reason is the following. The OPE of the
LHS of Eq.(\ref{KMN}) means that  it is obtained from
the short-distance expansion of the correlator; namely in the
large $-q^2$ limit.  Also being the OPE parts, $\Pi_N^{OPE}(p)$
and $\Pi_{K}^{OPE}(p-q)$ are obtained in the large $-p^2$ and
$-(p-q)^2$ limit respectively.   However,
there are other important regions of $p^2$ which contribute to
the OPE of the LHS.   An example of such regions are
given in table~\ref{table1}.
\begin{table}
\centering
\begin{tabular}{cccc}
\hline
contribution to $\Pi^{2HR,OPE}(q)$ & $\Pi_N^{OPE} (p)$ & 
$\Pi_K^{OPE} (p-q)$ & comments \\%[1ex]
\hline
%{} &{} &{} &{} \\%[-1.5ex]
large $-q^2$ & large $-p^2$ & large $-(p-q)^2$ & in Eq.(\ref{KMN}) \\%[1ex]
large $-q^2$ & small $-p^2$ & large $-(p-q)^2$ & not in Eq.(\ref{KMN}) \\%[1ex]
large $-q^2$ & large $-p^2$ & small $-(p-q)^2$ & not in Eq.(\ref{KMN}) \\%[1ex]
\hline
\end{tabular}
\caption{Typical momentum regions which contribute to the OPE of
$\Pi^{2HR}(q)$.  The first line represents the region which has
been taken into account through Eq.(\ref{KMN}). }
\label{table1}
\end{table}
Another serious problem with Eq.(\ref{KMN}) is the implicit
assumption of
\begin{eqnarray}
\langle 0| J_K J_N | K N \rangle =\langle 0| J_K | K \rangle
\times \langle 0| J_N | N \rangle,
\end{eqnarray}
which can be shown to be not true in general.

\subsection{Our method}

Here, we suggest to subtract out the 2HR contribution by
explicitly estimating the contribution coming from the
non-interacting K-N intermediate state,
\begin{eqnarray}
\Pi^{2HR}(q) & = & i|\lambda_{KN}|^2 \int \frac{d^4 p} {(2 \pi)^4}
\frac { \gamma_5 (\fslash{p}+m_N ) \gamma_5}{p^2 -m_N^2 } \frac{1}
{(p-q)^2-m_K^2} \label{phen2}
\end{eqnarray}
where
\begin{eqnarray}
\langle 0| J_\Theta|KN(p) \rangle= \lambda_{KN} i\gamma_5  u_N(p) \
. \label{lambdakn}
\end{eqnarray}
There are additional contributions coming from excited kaon or
nucleon states.   However, these contributions are exponentially
suppressed after the Borel transformation.   Hence, to estimate the
lowest 2HR contribution, we need to know the overlap strength in
Eq.(\ref{lambdakn}). This strength will be estimated in the
following section by combining the soft-kaon limit and a sum rule
for the nucleon with pentaquark interpolating field.

\section{Estimating the overlap strength}

To calculate the overlap strength $\lambda_{KN}$, we first use the
soft-kaon theorem,
\begin{eqnarray}
\langle 0| J_\Theta|KN \rangle
\stackrel{soft-kaon}{\longrightarrow} -\frac{1}{f_K} \langle 0|
[Q_5^K,J_\Theta]|N \rangle
& = & -\frac{1}{f_K} \langle 0| J_{N,5}|N \rangle \nonumber \\
& = & -\frac{\lambda_N}{f_K} i \gamma_5 u_N(p)  \label{softk}
\end{eqnarray}
where $Q_5^K=\int d^3 y d^\dagger(y) i \gamma_5 s(y)$ and
\begin{eqnarray}
J_{N,5} & = & \epsilon^{abc} \epsilon^{def} \epsilon^{cfg} \bigg[
\{ u^T_a C \gamma_5 s_b \} \{ u^T_d C \gamma_5 d_e \} C\bar{s}^T_g
+ \{ u^T_a C d_b \} \{ u^T_d C  s_e \}  C\bar{s}^T_g \nonumber \\
&& + \{ u^T_a C d_b \} \{ u^T_d C \gamma_5 d_e \}  C \gamma_5
\bar{d}^T_g \bigg]\ . \label{jn5-2}
\end{eqnarray}
Using Eq.(\ref{softk}) in Eq.(\ref{lambdakn}) we have,
\begin{eqnarray}
\lambda_{KN} & = & - \frac{1}{f_K} \lambda_N\ .
\label{jn5-1}
\end{eqnarray}
Hence, to estimate $\lambda_{KN}$, we need to calculate $\lambda_N$
which represents  the five-quark
component of the nucleon.

To do that,
we first construct the sum rule for the nucleon using the
following ``old-fashioned'' correlation function,
\begin{eqnarray}
\Pi(q)=i \int d^4 x e^{iq\cdot x} \langle 0 | \theta(x^0) J_{N,5}(x)
\bar{J}_{N,5} (x) |0 \rangle\ , \label{corr2}
\end{eqnarray}
where $J_{N,5}$ is given in Eq.(\ref{jn5-2}).
This type of ``old-fashioned''
correlation function has been successfully used
in projecting out
positive and negative parity nucleon~\cite{Jido:1996ia}.
We then divide the
imaginary part into two parts for $q_0>0$,
\begin{eqnarray}
\frac{1}{\pi} {\rm Im} \Pi(q_0)=A(q_0) \gamma^0+B(q_0)\ .
\end{eqnarray}
Then the spectral density for the positive and negative parity
physical states will be as follows,
\begin{eqnarray}
\rho^{\pm}(q_0)=A(q_0)\mp B(q_0)\ .
\end{eqnarray}
Note that the signs are reversed compared to that of SDO because
the nucleon current $J_{N,5}$ as given in Eq.(\ref{jn5-2}) has an
additional factor of $\gamma_5$ compared to the usual nucleon
current.

\begin{figure}[ht]
%\epsfxsize=10cm   %width of figure - will enlarge/reduce the figures
%\epsfbox{fig3.eps}
%\figurebox{2cm}{3cm}{} %to have a box alone
\centerline{\epsfxsize=5.1in\epsfbox{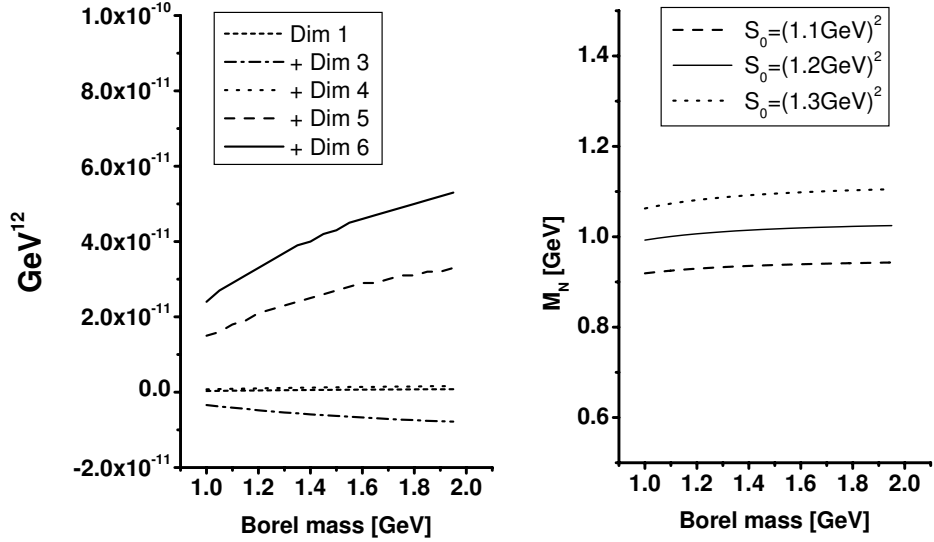}}
\caption{The figure in the left panel shows the Borel
curves for $|\lambda_{N+}|^2 e^{-m^2_{N+} / M^2}$ from
Eq.~(\ref{nucsum}) as we add up the OPE order by order.
The figure in the right panel shows the mass of the nucleon from the five-quark
current with specified continuum threshold. \label{inter}}
\end{figure}

For the nucleon correlation function given in Eq.(\ref{corr2}), the
respective OPE are given by
\begin{eqnarray}
A_{OPE}(q_0) & =& \frac {3 q_0^{11}} {5! 5! 2^{10} 7 \pi^8} +
\frac {4 q_0^{7} m_s \langle \bar{s} s \rangle} {3! 5! 2^8 \pi^6}
+\frac{3 q_0^7}{3! 5! 2^{10} \pi^6}\left\langle \frac{\alpha_s}{\pi} G^2
\right\rangle
%\nonumber \\
%&&
- \frac {4 q_0^{5}} {3! 4! 2^9 \pi^6} m_s \langle \bar{s} g \sigma \cdot
G
s \rangle
%\nonumber
\\
B_{OPE}(q_0) & =& \frac {2 q_0^{10} m_s } { 5! 5! 2^{10} \pi^8}
- \frac { q_0^{8} \left[ 2 \langle \bar{s} s \rangle
-\langle \bar{d} d \rangle \right ] } { 4! 5! 2^7 \pi^6} 
%\left[ 2 \langle \bar{s} s \rangle
%-\langle \bar{d} d \rangle \right ]
%\nonumber \\
 + \frac {q_0^{6}} { 3! 4! 2^9 \pi^6} \left [ 2 \langle \bar{s} g
\sigma \cdot G s \rangle - \langle \bar{d} g \sigma \cdot G d
\rangle \right ] \ .
\end{eqnarray}
Here, we have kept the extra numeric factors in the numerator so that our 
OPE can be directly compared with that of the SDO sum rule. From 
the comparison, we see that the dimension-even operators have been amplified 
by the factor 3 or 4 while for the dimension-odd operators 
(dimension 3 and 5) there are
partial cancellations among them.

The spectral density is assumed to have the following form,
\begin{eqnarray}
\rho_{phen}^\pm(q_0)=|\lambda_{N\pm}|^2 \delta(q_0-m_{N\pm}) +\theta(q_0
-\sqrt{s_0}) \rho^\pm_{cont}(q_0)\ ,
\end{eqnarray}
where the usual duality assumption has been used to represent
the higher resonance contribution above the continuum threshold
$\sqrt{s_0}$.
We substitute this into the following Borel transformed dispersion
relation,
\begin{eqnarray}
\int^\infty_0 dq_0 e^{-q_0^2/M^2 } [\rho_{phen}^\pm(q_0)-\rho_{OPE}^\pm (q_0)]
 = 0\ ,
\end{eqnarray}
and obtain a sum rule for $|\lambda_{N\pm}|^2$
\begin{eqnarray}
|\lambda_{N\pm}|^2 e^{-m_{N\pm}^2/M^2 } = \int_0^{\sqrt{s_0}}
dq_0~ e^{-q_0^2/M^2} \rho_{OPE}^\pm (q_0)\ .
\label{nucsum}
\end{eqnarray}
The sum rule for
the nucleon mass is obtained by taking the derivative with respect to $1/M^2$.
Using the same QCD parameters as in Ref.\cite{Oka1}, we plot
the RHS of Eq.(\ref{nucsum}) and the Borel curve for the nucleon mass
in Fig.\ref{inter}.
As can be seen from the figure in the left panel, the OPE with
dimension 5, which contains the quark-gluon mixed condensate,
still contributes to the sum rule appreciably. This feature is similar
to the SDO sum rule where the quark-gluon mixed condensate is
the main origin for yielding the negative-parity. 
However, in our case, we have additionally 
important contribution from the dimension 6 operator. 
Because of this, 
the pentaquark nucleon current does not exclusively couple to a
specific parity state but it couples to both parities.  
In fact, from Fig.\ref{inter},
we obtain a consistent (positive) sum rule for
$|\lambda_{N+}|^2 \sim 1 \times 10^{-10}~{\rm GeV}^{12}$ and a reasonable
mass for the nucleon. Similarly we also obtain consistent results
for the negative-parity nucleon $S_{11}(1535)$ from
the sum rule for $|\lambda_-|^2$.

\section{Reanalysis of SDO sum rule}

\begin{figure}[ht]
%\epsfxsize=10cm   %width of figure - will enlarge/reduce the figures
%\epsfbox{fig3.eps}
%\figurebox{2cm}{3cm}{} %to have a box alone
\centerline{\epsfxsize=5.1in\epsfbox{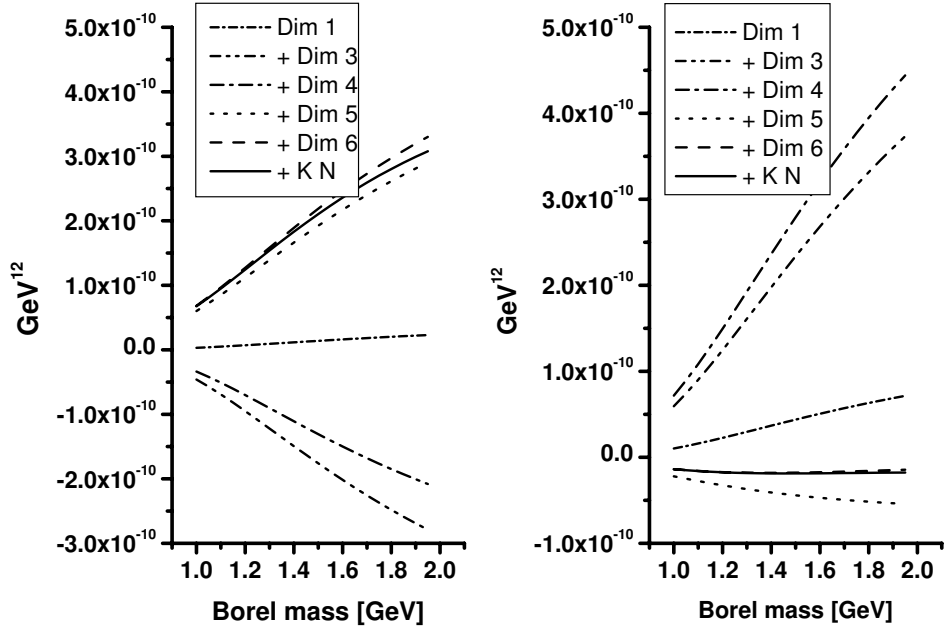}} \caption{Borel
curves for the LHS of Eq.~(\ref{fsum}) for the negative-parity
(left panel) and positive-parity (right panel) pentaquark as we
add up the OPE order by order. The solid lines are
obtained when the K-N 2HR contribution is subtracted from the OPE.
\label{newoka}}
\end{figure}

We now reanalyze the SDO sum rule including the K-N 2HR contribution
given by Eq.(\ref{phen2}).
The imaginary parts of Eq.(\ref{phen2}) for the positive and negative parity
channels are
\begin{eqnarray}
\rho_{KN}^\pm(q_0)& = & \frac{|\lambda_{KN}|^2}{32 \pi^2 } \sqrt{
(q_0-m_K)^2-m_N^2 } \sqrt{(q_0+m_K)^2-m_N^2 } \nonumber
\\ && \times
\frac {(q_0\mp m_N)^2-m_K^2} {q_0^3}\ .
\end{eqnarray}
Here we use $|\lambda_{KN}|^2= |\frac{1}{f_K} \lambda_{N+}|^2$
as determined from the previous section.
Then, the SDO sum rule with the explicit contribution from the
K-N 2HR contribution subtracted out reads
\begin{eqnarray}
|\lambda_{\Theta\pm}|^2 e^{-\frac{M_{\pm}^2} {M^2}} =
\int\limits_0^{\sqrt{s_0}} dq_0~ e^{-\frac{q_0^2}{M^2}}
\rho^{\pm}_{OPE}(q_0) -\int\limits_{m_K+m_N}^\infty
dq_0~e^{-\frac{q_0^2}{M^2}} \rho_{KN}^\pm(q_0) \label{fsum}
\end{eqnarray}
where $\rho^{\pm}_{OPE}$ is given in Ref.\cite{Oka1}.
Note, the LHS is positive definite. The reliable sum rule should
have the RHS with the same sign.
The RHS of Eq.~(\ref{fsum}), which includes
the OPE as well as the K-N 2HR contribution, is
plotted in Fig.\ref{newoka}.  We see that for the negative-parity
case the RHS is positive agreeing with the sign in the LHS.
But for the positive-parity case, the sign of the RHS does not
satisfy the constraint on sign required by the LHS.
As can be seen also in the figure, the contribution from the 2HR state
constitutes less than 10\% of the total OPE.
Hence, the conclusion first given by SDO
that the OPE is consistent with the existence of a negative parity
pentaquark state remains valid.

In summary, we have reanalyzed the QCD sum rule for $\Theta^+$
with the K-N 2HR contribution being subtracted out.
The strength for the K-N 2HR with the $\Theta^+$ interpolating
field has been estimated by the soft-kaon theorem and the resulting
nucleon sum rule with five-quark current.
The K-N continuum contribution was found to be less than 10 \% and
the QCD sum rule for $\Theta^+$ still supports a negative-parity state.

\acknowledgments

This work was supported by 
Korea Research Foundation Grant (KRF-2004-015-C00116).

\end{document}